\documentclass[aps,reprint,amsmath,amssymb,superscriptaddress]{revtex4-2}
\usepackage{graphicx}
\usepackage{hyperref}
\usepackage{xcolor}
\usepackage{todonotes}

\begin{document}
\title{Unveiling universal aspects of the cellular anatomy of the brain}

\author{Helen S. Ansell}
\affiliation{Department of Physics and Astronomy, Northwestern University, Evanston, Illinois 60208, USA}

\author{Istv\'{a}n A. Kov\'{a}cs} 
\email{Corresponding author: istvan.kovacs@northwestern.edu}
\affiliation{Department of Physics and Astronomy, Northwestern University, Evanston, Illinois 60208, USA}
\affiliation{Northwestern Institute on Complex Systems, Northwestern University, Evanston, Illinois 60208, USA}
\affiliation{Department of Engineering Sciences and Applied Mathematics, Northwestern University, Evanston, Illinois 60208, USA}

\begin{abstract}
Recent cellular-level volumetric brain reconstructions have revealed high levels of anatomic complexity. Determining which structural aspects of the brain to focus on, especially when comparing with computational models and other organisms, remains a major challenge. Here we quantify aspects of this complexity and show evidence that brain anatomy satisfies universal scaling laws, establishing the notion of structural criticality in the cellular structure of the brain. Our framework builds upon understanding of critical systems to provide clear guidance in selecting informative structural properties of brain anatomy. As an illustration, we obtain estimates for critical exponents in the human, mouse and fruit fly brains and show that they are consistent between organisms, to the extent that data limitations allow. Such universal quantities are robust to many of the microscopic details of individual brains, providing a key step towards generative computational brain models, and also clarifying in which sense one animal may be a suitable anatomic model for another.
\end{abstract}

\date{\today}

\maketitle

\section{Introduction}
Mapping out the anatomic and connectomic structures of the brain is a current large-scale effort in neuroscience that will give unique insights into the structure and function of the brain~\cite{human,mouse,fly,mouse23,DeWeerdt2019,Celegans,Witvliet2021,tadpole,PdumeriliiLarva,flylarva,zebrafishlarva,flywire,Zheng2018, mousemind}. 
For small nervous systems, there are already complete nanoscale reconstructions, including the nematode \textit{Caenorhabditis elegans}~\cite{Celegans,Witvliet2021}, the tadpole larva of \textit{Ciona intestinalis}~\cite{tadpole}, the marine annelid \textit{Platynereis dumerilii}~\cite{PdumeriliiLarva}, and the fruit fly (\textit{Drosophila melanogaster}) larva brain~\cite{flylarva}.
For larger brains, partial volumetric reconstructions are available in the human~\cite{human}, mouse~\cite{mouse,mouse23}, fruit fly~\cite{fly}, and zebrafish larva~\cite{zebrafishlarva} with data from larger volumes, including the full fruit fly brain~\cite{flywire,Zheng2018}, and additional organisms expected over the next few years~\cite{DeWeerdt2019}.

These detailed brain reconstructions open up the possibility of making structural comparisons between the brains of different organisms.
However, meaningful comparisons require analysis of largescale data.
The largest currently available reconstructions, consisting of \(1\,\mathrm{mm}^3\) of human~\cite{human} and mouse~\cite{mouse} brain, each require over a petabyte (1 petabyte = 1000 terabytes) of storage, while future whole brain mappings will be orders of magnitude larger~\cite{DeWeerdt2019,mousemind}.
Making full use of all this information not only poses big data and machine learning challenges, but also requires decisions on which aspects of the anatomic structure to focus.

Identifying overarching structural features in the complexity of the brain can aid in understanding and modeling its structural properties and their relation to function. 
At the cellular level, complexity in the structure of neurons has frequently been quantified through the fractal dimension $d_f$.
Typical neuron \(d_f\) values are between 1.1-1.9~\cite{Cross1994,Takeda1992,Alves1996,Smith2021}, with variation between organisms, neuron types and developmental stage, as well as details of the methodology~\cite{Smith1996,Jelinek2005}. 
It has been hypothesized that $d_f$ in neurons is related to the 
trade-off between connectivity and wiring cost for a neuron, with higher connectivity requirements resulting in higher $d_f$~\cite{Smith2021}.

Fractal behavior is an example of scale invariance, or self-similarity in the structure of the brain.
Self-similar behavior has been widely reported in both the structure and function of the brain at a range of scales, as reviewed in Ref.~\cite{Grosu2022}.
For example, self-similarity in the macroscale structure is observed in the gyrification of the cerebral cortex~\cite{Mota2015}.
At the microscale, self-similarity is present in the dendritic branching of individual neurons, as can be detected through measuring correlations in the structures~\cite{Alves1996,Wen2009}, and box-counting techniques~\cite{Takeda1992,Alves1996, Smith2021}.
The relationship between these self-similar spatial features and scale invariant functional properties remains not well understood~\cite{Grosu2022, Batista2018}.
Deeper understanding of the structure of the brain will aid further exploration of this relationship.

Here, we propose that statistical physics can provide a guiding framework for determining and quantifying additional structural features in the cellular complexity of the brain.
By analyzing properties related to cell size, as well as pairwise and higher-order correlations in volumetric brain regions, we show that the cellular structure of the brain displays signatures of collective behavior close to criticality.
These features include self-similarity in the size-distribution of cell fragments within sample regions, and long-range spatial correlations in the structure.
From these measurements, we estimate a set of exponents in each brain.
We find that these exponents obey critical scaling relations, further indicating that the brain is in the vicinity of \textit{structural criticality}, which is distinct from observations of critical behavior in neural dynamics~\cite{Beggs2003}.
The relations between critical exponents mean that these different structural properties are not independent in the brain, but are different manifestations of the same emergent phenomena. 
These exponents therefore form a structural universality class in each brain.

We also observe that the critical exponents are compatible between organisms. 
This suggests that key structural properties of the brains of a broad range of organisms may be guided by similar principles that can be quantified through these universal properties.
If such properties of the cellular structure of the brain are indeed universal, the compatibility between organisms can be described in terms of a structural \textit{brain universality class}.
Studying simple generative models within the brain universality class opens up the possibility of generating spatial structures that capture the emergent large-scale properties of cellular brain anatomy. 
Such models could be studied to gain insight into the relationship between brain structure and function.
The exponents can also act as a baseline for comparison between structural properties of spatial models with brain-like features, such as models of physical networks~\cite{Pete2023}, and the brain.

\begin{figure}
\centering
\includegraphics[width=86mm]{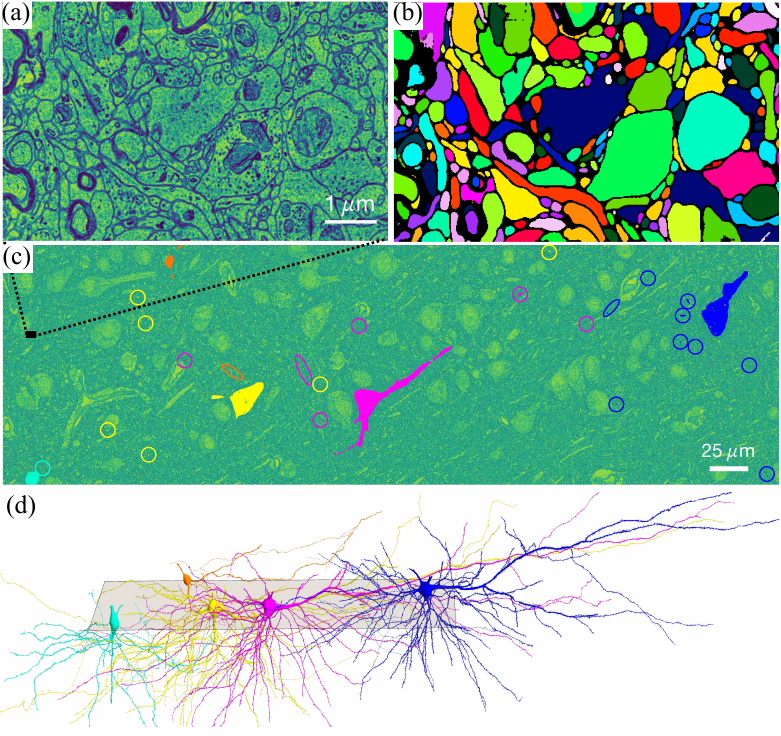}
\vskip-5mm
\caption{Illustration of data used from the human brain dataset~\cite{human}.
(a) EM image of a region of the human brain data.
(b) Segmentation data for the EM image in (a). Colors indicate distinct segments. 
(c) EM image of a larger region of the human brain with the segmentation for five select neurons highlighted, showing how the same neurons may cross a given plane multiple times. Circles are used to highlight the locations of small fragments that are not easily visible. The black rectangle in the upper left indicates the region shown in (a).
(d) 3D mesh reconstructions of the five neurons highlighted in (c) with the larger area shown in (c) indicated by the gray rectangle.}
\label{fig:datasets}
\end{figure}

\section{Sampling brain datasets}
We use publicly available volumetric brain datasets consisting of \(\sim 1\,\mathrm{mm}^3\) each of female human~\cite{human} and male mouse~\cite{mouse} cortical tissue, as well as half of a female fruit fly (\textit{Drosophila melanogaster}) central brain~\cite{fly}.
These datasets are large enough in scale that our statistical sampling techniques can be conclusively applied, which would not be possible on the smaller nervous system organisms for which full volumetric reconstructions are currently available~\cite{Celegans,Witvliet2021,tadpole,PdumeriliiLarva,flylarva}.

Each dataset consists of a brain volume that was sliced and imaged using electron microscopy (Fig.~\ref{fig:datasets}(a), (c)) before the individual cells were identified and reconstructed (Fig.~\ref{fig:datasets}(b--d)) to recreate the full 3D volume.
While in the fly, the neurons are all fully proofread~\cite{fly}, the human~\cite{human} and mouse~\cite{mouse} datasets have currently undergone partial cell-type identification and proofreading. As such, we primarily present our analysis including all cell types, although we show that including only neurons gives qualitatively similar results. 

Here, we randomly sample volumetric regions of the segmentation data, which identifies the cell present at each voxel location, at a voxel side length of 128 nm.
These samples contain positional information about the cells present within 3D volumes and 2D slices of the segmentation data, as shown in the schematics in Figs. ~\ref{fig:size-distribution}(a) and~\ref{fig:correlations}(a) respectively.
The cell fragments, or segments, in the samples have the appearance of clusters on a simple cubic lattice, with the lattice spacing determined by the sampling resolution.
We generate sets of at least 1000 samples of linear size \(L\) in each brain, and study finite-size effects by sampling \(16\leq L \leq 512\) voxels in 3D. 
In 2D, we sample up to \(L = 4096\) voxels in the mouse and human, and \(L = 1024\) voxels in the fly.
Details of the sampling, including choice of resolution, are discussed in Methods.

\section{Estimating critical exponents}
\begin{figure*}
\centering
\includegraphics[width=178mm]{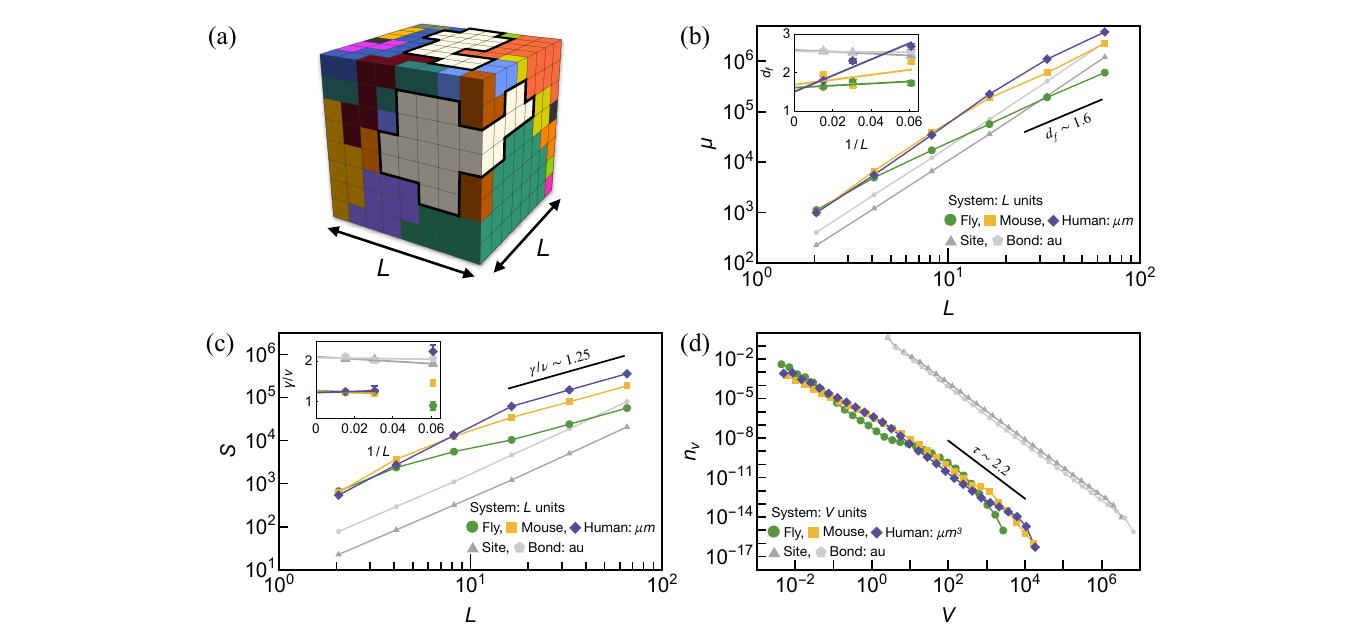}
\vskip-5mm
\caption{Analysis of segment size properties in 3D brain samples.
(a) Schematic of a 3D brain sample of size \(L = 8\), with colors indicating individual segments. The largest mass segment (cream) is outlined in black.
(b) \(L\)-dependence of the mean largest segment size \(\mu\) in the 3D samples, plotted on logarithmic axes. Results in each plot are also shown for critical site and bond percolation, which are known to display structural criticality.
(inset) FSS of the fractal dimension \(d_f\) extracted from \(\mu(L)\) as described in Methods.
(c) \(L\)-dependence of the mean segment size \(S\) in each sample, plotted on logarithmic axes. 
(inset) FSS of the exponent \(\gamma/\nu\) (discussed in Methods).
(d) Size distribution of segments in the largest 3D sample sizes with points logarithmically binned by volume. 
Black lines in (b--d) indicate the mean exponent value for the brains.
Error bars in the main panel in (b) and (c), indicating the standard error on the mean, are smaller than the point size.
Error bars in inset figures are propagated uncertainties from points in the corresponding main figure panel.
}
\label{fig:size-distribution}
\end{figure*}

In models of clustered systems, information about structural properties of the system, such as phase behavior, is encoded in the behavior of properties related to the size, or mass, of clusters as well as correlations in the structure. 
Close to criticality, such properties are well-approximated by power-law relations.
We investigate these quantities in the generated samples from the brain segmentation data to gain insight into structural properties of the brain.

We first investigate the mean size \(\mu\) of the largest segment in the 3D brain samples, plotted in Fig.~\ref{fig:size-distribution}(b).
We observe that \(\mu(L)\) appears consistent with the relation \(\mu(L)\sim L^{d_f}\), which is well-understood as the form \(\mu(L)\) takes at criticality in percolation~\cite{Wang2013}, and  demonstrates the expected presence of fractal-like behavior in the brain samples.
The slope of the \(\mu\) vs \(L\) curve is influenced by finite-size effects. 
In a critical system, the relation \(\mu(L)\sim L^{d_f}\) is expected to hold at sufficiently large $L$.
To estimate the exponent $d_f$ in the brains, we use finite-size scaling (FSS) techniques to extrapolate the large $L$ behavior, as is well understood in physical systems~\cite{CardyFSS} and has also been applied to a range of biological systems~\cite{Klaus2011,Attanasi2014,Corral2016, Tang2017,Martin2021}.
We use two-point fits between points at successive $L$ values to estimate size-dependent \(d_f\) values. 
We then plot these values as a function of $1/L$ and use a linear fit to extrapolate $d_f$ in the large $L$ limit ($1/L\to 0$), as shown in the inset of Fig.~\ref{fig:size-distribution}(b), and discussed in more detail in Methods. 
The obtained \(d_f\) values, listed in Tab.~\ref{tab:critical-exponents}, are consistent with \(d_f\sim 1.6\) in all three organisms.
To provide a benchmark with a system known to display structural criticality, we also plot results for 3D critical site and bond percolation (discussed in Methods).
By matching the number of brain samples, the percolation uncertainty can set an independent expectation on the precision of the brain results.

\begin{figure*}
\centering
\includegraphics[width=178mm] {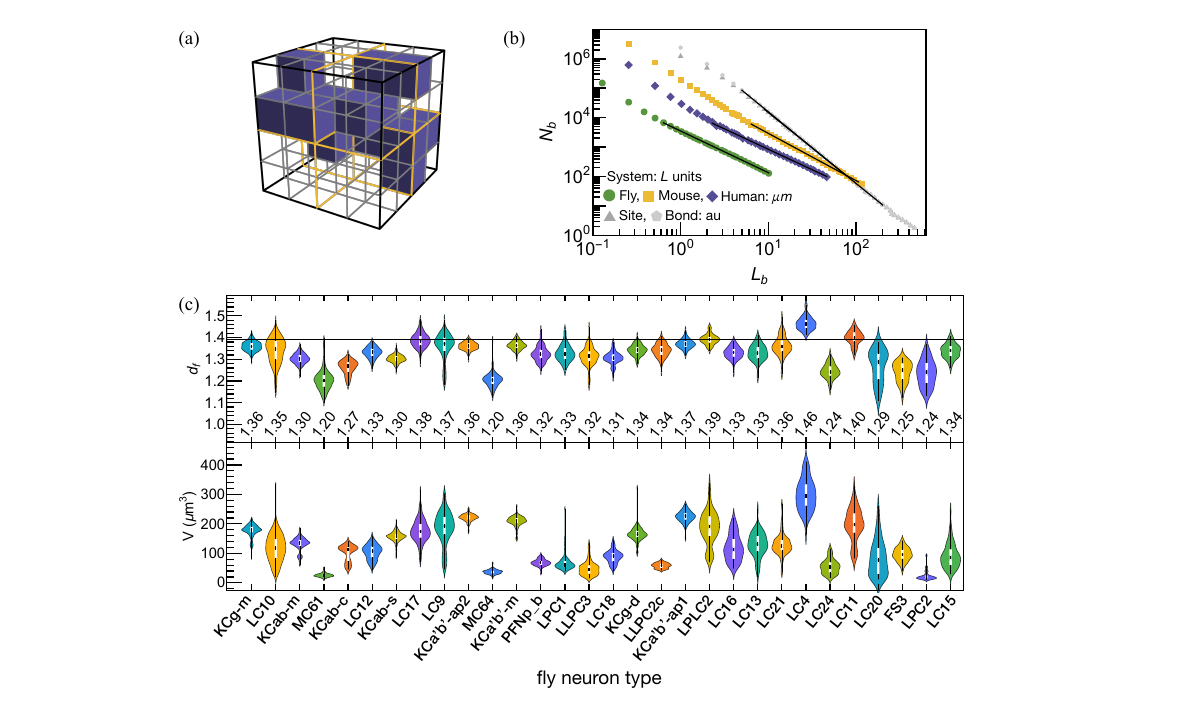}
\vskip-5mm
\caption{
Neuron box fractal dimension calculation.
(a) Schematic of grids of different sizes used in the box-counting procedure. Shown are \(L_b\) = 1 (gray), \(L_b = 2\) (yellow), and \(L_b = 4\) (black).
(b) Log-log plot of \(N_b\) vs \(L_b\) for 3D neuron reconstructions and the largest cluster in the largest size percolation samples. The magnitude of the best fit slope gives the \(d_f^{(box)}\) values listed in Tab.~\ref{tab:critical-exponents}. The standard error on each point is smaller than the point size.
(c) Distribution plots of \(d_f^{(box)}\) values (top) and volume (bottom) for the 30 most frequently occurring neuron types in the fly hemibrain volume~\cite{fly}.
In the top panel, median \(d_f^{(box)}\) values are indicated below each distribution and the horizontal black line gives the fly \(d_f^{(box)}\) value calculated in (b).
The median of each distribution is marked in black, while the interquartile range is indicated in white. 
Note that while the neurons included are classified as ``uncropped'' in the dataset, parts of the neuron, including the soma, may be missing from the data, which could lead to underestimation of the total volume.
}
\label{fig:fly-neurons}
\end{figure*}

We further investigate fractal properties of select neurons in the samples using the frequently-used box counting method. 
This involves overlaying an object with a grid of side length \(L_b\) and counting the number of grid cells \(N_b\) containing part of the object (see Fig.~\ref{fig:fly-neurons}(a) and Methods). 
The box-counting fractal dimension \(d_f^{(box)}\) is defined by
\begin{equation}
	d_f^{(box)} = \lim_{L_b\to 0} \frac{\log{N_b(L_b)}}{\log{1/L_b}}.
\end{equation}	
In practice $d_f$ is determined from the slope of the linear region of a log-log plot of $N_b$ vs $L_b$, which is also the region over which the structure displays self-similarity.
Here, we perform 3D box-counting on reconstructions of select neurons (discussed in Methods), with averaged results for each brain shown in Fig.~\ref{fig:fly-neurons}(b) and Tab.~\ref{tab:critical-exponents}.
In Fig.~\ref{fig:fly-neurons}(c), we plot distributions of \(d_f^{(box)}\) for the 30 most frequent neuron types in the fly dataset~\cite{fly}, which shows that the mean \(d_f^{(box)}\) value is representative of a ``typical'' fly neuron despite noticeable differences between some neuron types.
We observe that \(d_f^{(box)}<d_f^{(\mu)}\) in the brains and percolation.
For percolation, where the calculations can be directly compared, \(d_f^{(\mu)}\) is much closer to the known \(d_f\) value than \(d_f^{(box)}\), indicating that box-counting appears less reliable than the FSS approach at these sizes and numbers of samples.

We now consider the mean segment size per voxel, \(S\), within the 3D samples. This is given by~\cite{Wang2013}
\begin{equation}
	S = \frac{1}{V}\sum_i m_i^2,
\end{equation}
where \(V\) is the sample volume, and the sum includes the masses \(m_i\) of all sample segments. 
At criticality, in a finite sample, this scales as \(S(L)\sim L^{\gamma/\nu}\) ~\cite{Wang2013}, where \(\gamma\) is the susceptibility exponent and \(\nu\) is the correlation length exponent.
For our 3D samples, Fig.~\ref{fig:size-distribution}(c) shows \(S(L)\), which indicates that the brain displays scaling behavior associated with criticality. 
The inset shows the FSS of the exponent \(\gamma/\nu\) (discussed in Methods).
The extrapolated brain exponent values, listed in Tab.~\ref{tab:critical-exponents} are compatible across the studied organisms and have mean value \(\gamma/\nu\sim1.25\).

A third property related to segment sizes is their distribution within the samples.
At criticality, scale invariance results in the entire system displaying self-similarity and a broad segment size distribution, in addition to the self-similarity of individual segments. 
Fig.~\ref{fig:size-distribution}(d) shows the segment size distribution in our largest 3D samples, with datapoints logarithmically binned.
The observed behavior shows
slow decay of this quantity over several orders of magnitude, as is expected in a system at criticality. 
This measurement, along with \(S\), therefore provides evidence beyond single-cell measures that the brain displays signatures of structural criticality.
The Fisher exponent \(\tau\) that characterizes this behavior is expected to be related to the previously calculated exponents through (see Methods)
\begin{equation}
	\tau = 3-\frac{\gamma/\nu}{d_f}.
	\label{eq:tau-gamma}
\end{equation}
Using this relation to estimate \(\tau\), we find that the values, listed in Tab.~\ref{tab:critical-exponents2}, are consistent between organisms with \(\tau\sim 2.2\).
This value is indicated in Fig.~\ref{fig:size-distribution}(d), and appears consistent with the data, suggesting that the scaling relation holds true.

\begin{figure*}
\centering
\includegraphics[width=178mm]{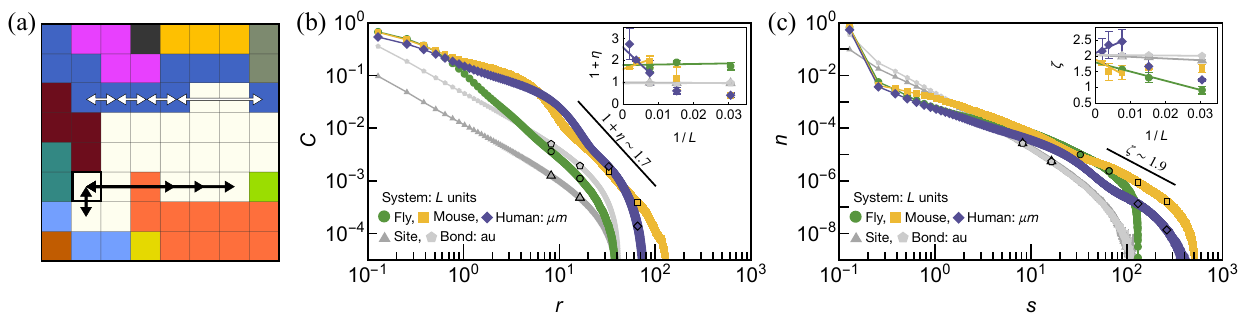}
\vskip-5mm
\caption{Long-range correlations in brain samples.
(a) Schematic of a 2D brain sample of size \(L=8\) voxels, where colors indicate different segments. Pairs of points that contribute positively to the pair correlation function \(C(r)\) starting from the highlighted point are indicated by black arrows, while white arrows indicate pairs of points in the blue cluster in the indicated row that contribute to the gap size statistics. 
(b) \(C(r)\) for the largest 2D sample size for each system on a log-log plot. Highlighted points are used in the two-point fit to determine \(\eta\) (see Methods).
(inset) FSS of the exponent of \(C(r)\sim r^{-d+2-\eta}\).
(c) Log-log plot of \(n(s)\) for the largest 2D sample size for each system. Highlighted points are used in the two-point fit to determine \(\zeta\). 
(inset) FSS of the \(\zeta\) exponent, \(n(s)\sim s^{-\zeta}\). 
Error bars in the main panel in (b) and (c) indicate the standard error, which is typically smaller than the point size, while the inset error bars are range bars indicating the values of the exponents at the two distances sampled in each case, as discussed in Methods. 
Black lines in (b) and (c) indicate the mean extrapolated exponent value for the brains. Note that in (b), the human value is treated as an outlier and is not included in the calculation of the mean value. 
}
\label{fig:correlations}
\end{figure*}


Going beyond size-related measures, we now check for the presence of long-range correlations in the brain, as a further indicator of criticality. 
We start with the connected two-point correlation function \(C(r)\), which compares the likelihood that two points at separation \(r\) belong to the same segment with the baseline assumption of a constant density.
This is defined by 
\begin{equation}
		C(|\mathbf{x}-\mathbf{x}'|) = \langle S(\mathbf{x})S(\mathbf{x}')\rangle-\langle S(\mathbf{x})\rangle\langle S(\mathbf{x}')\rangle, 
	\end{equation}
where \(r = |\mathbf{x}-\mathbf{x}'|\) is the separation between points at positions \(\mathbf{x}\) and \(\mathbf{x}'\), \(\langle\dots\rangle\) denotes averaging across possible locations, and \(S(\mathbf{x})\) has the form
	\begin{equation}
		S(\mathbf{x}) =     \begin{cases}
            1 \text{ if the site belongs to the segment}\\
            0 \text{ otherwise.}
       		\end{cases}
        \end{equation}
We calculate \(C(r)\) in the $x$ and $y$ directions for each segment in a 2D sample, and average across the two directions (see schematic Fig.~\ref{fig:correlations}(a)).
We do not calculate $C(r)$ in the $z$ direction due to different voxel dimensions in the human and mouse datasets and the smaller $z$ extent of each sample limiting attainable distances. 
While directionality is known to be important in the brain, we observe that our calculations are qualitatively similar along the $x$ and $y$ directions. We therefore average the two, treating the structure as isotropic to first approximation.
While the lack of periodic boundary conditions in our samples means that technically \(\langle S(\mathbf{x})\rangle\neq \langle S(\mathbf{x}')\rangle\), for simplicity we assume the segment distribution is such that \(\langle S(x) \rangle =\langle S(x') \rangle = \langle \rho\rangle\), the average density of the segment in the sample.

Fig.~\ref{fig:correlations}(b) shows $C(r)$ for the largest $L$ samples, while plots for all \(L\) are shown in Fig.~S1.
We observe that long-range correlations are indeed present in the brain, as evidenced by \(C(r)\) decaying slowly over more than two orders of magnitude.
At criticality, \(C(r)\sim r^{-d+2-\eta}\), where \(d=3\) is the spatial dimension and \(\eta\) is the anomalous dimension. 
Examining the FSS of the exponent (see Fig.~\ref{fig:correlations}(b, inset) and Methods) leads to the extrapolated \(\eta\) values in Tab.~\ref{tab:critical-exponents}, which are consistent between the fly and mouse with \(\eta\sim 0.7\).
While the human \(\eta\) value appears larger, the difference is small compared to the error.

The presence of higher-order long-range correlations can offer a deeper test of criticality. 
We probe higher-order correlations by examining the ``gap-size statistics'' \(n(s)\). 
A pair of sites in a segment at separation \(s\) contributes to \(n(s)\) only if there are no other sites belonging to the same segment
between the pair, as indicated in Fig.~\ref{fig:correlations}(a). 
This contrasts \(C(r)\), where all pairs of sites in a segment contribute, and leads to \(n(s)\) capturing higher-order structural complexity related to the concavity of the segment shape.
Fig.~\ref{fig:correlations}(c) shows \(n(s)\) for the largest \(L\) 2D samples (all \(L\) are shown in Fig.~S2), in which the slow decay of \(n(s)\) over several orders of magnitude indicates long-range correlations.
The FSS of the exponent \(n(s)\sim s^{-\zeta}\) (Fig.~\ref{fig:correlations}(c, inset), also see Methods) leads to the extrapolated \(\zeta\) values in Tab.~\ref{tab:critical-exponents}.
In the mouse and human \(\zeta\sim 2\), while in the fly \(\zeta\) is slightly smaller. 
However, the datapoint trend suggests that with larger sizes available \(\zeta\) could converge towards 2.
This is in line with known numerical or analytic $\zeta$ values in other critical systems, which all indicate $\zeta=2$ ~\cite{Kovacs2012gap,Kovacs2014Potts,Kovacs2014Perc},
further implying critical behavior in the brain.

\section{Analyzing neurons}
While our analysis so far has mostly included all cell types, for many applications the physical structure of neurons may be of particular interest. 
Restricting the analysis to only neuron segments (discussed in Methods), we find that neurons qualitatively agree with our previous results. This agrees with the expectation that large-scale behavior is dominated by neurons as other cell types have a much more limited spatial extent.
We quantify the neuron behavior in Tab.~S1, which lists exponent values for neurons and all segments at the largest \(L\) available in all three brains, and observe good agreement between datapoints.

One set of points that do not overlap within the errors are between the \(\zeta\) values for the human sample. 
At the largest $L$ available in the human, we find $\zeta(L) = 2.2(4)$ for all segments and 1.9(6) for the neuron segments, suggesting the large $L$ behavior is consistent between neurons and all segment types. 
The discrepancy at smaller $L$ may be due to this being the size above which correlations in the human brain can be considered long-range.

The other non-overlapping errors are between the \(d_f^{(\mu)}\) values in the mouse for neurons and all segments, and between the \(\zeta\) values in the mouse for neurons at different resolutions. 
We attribute these to the lower proportion of currently available cell-type identification in this dataset (see Fig.~S3), which is also what limits the statistics for obtaining good estimates for the extrapolated exponents for neurons in the mouse and human.
This limitation is due to the current level of proofreading in the datasets and could be revisited after more extensive cell identification has been undertaken. 
Overall, these results suggest that 
analysis including all cell types
is a good approximation for the behavior of neurons at the largest scales. 

\section{Testing for proximity to criticality}
We have shown that the brain displays signatures of structural criticality, but require further analysis to determine whether the structure is indeed critical.
One consideration is whether the structure is at criticality, or slightly off-critical.
While typically the critical point can be determined by tuning the control parameter, the lack of accessible control parameter here necessitates an alternative approach.

\begin{figure}
\centering
\includegraphics[width=86mm]{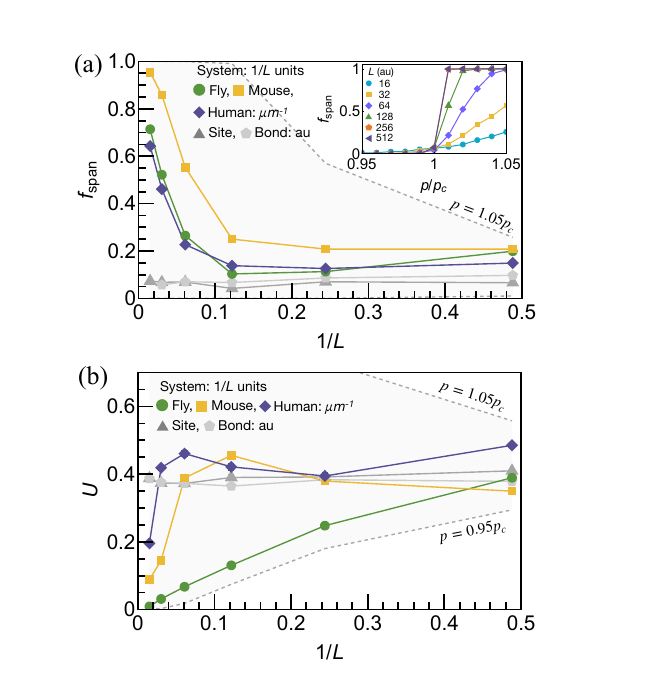}
\vskip-5mm
\caption{Testing for the proximity to criticality.
(a) Fraction of 3D samples containing at least one spanning cluster \(f_\mathrm{span}\) in samples of size \(L\). 
(inset) Spanning fraction for site percolation with different occupation probabilities \(p/p_c\), where $p_c$ is the critical probability, from within the shaded area in the main panel for different \(L\) values.
(b) Binder cumulant \(U\) (Eq.~\ref{eq:Binder}) calculated for each brain and critical percolation for different \(L\) values. 
In both panels, the shaded region highlights the numerically-obtained region in which site percolation points within 5\% of \(p_c\) fall.
}
\label{fig:criticality-tests}
\end{figure}

Motivated by percolation theory, one possible test of criticality is to examine the proportion of samples containing a ``spanning'' segment that touches all faces of the bounding cube and is connected within the sample.
While spanning fraction is an order parameter of percolating clusters, and can be calculated for brain samples, the nature of the structural brain order parameter remains an open question.
Fig.~\ref{fig:criticality-tests}(a) shows the fraction of 3D samples containing a spanning segment, which is observed to increase with \(L\).
This behavior is consistent with the trend of percolation slightly in the ordered phase (Fig.~\ref{fig:criticality-tests}(a, inset)), where the behavior is dominated by a large cluster.
If spanning is a meaningful order parameter for the brain, this could indicate that the brain is slightly in the ordered phase.

Another quantity we can estimate is the fourth-order Binder cumulant~\cite{Binder1981}, 
	\begin{equation}
		U = \frac{3}{2}\left(1-\frac{\left<\mathcal{M}^4\right>}{3\left<\mathcal{M}^2\right>^2}\right),
		\label{eq:Binder}
	\end{equation}
with~\cite{Ballesteros1997}
	\begin{align}
		\left<\mathcal{M}^2\right> &= \frac{1}{V^2}\sum_i m_i^2 = \frac{S}{V}\\		
		\left<\mathcal{M}^4\right> &= 3\left<\mathcal{M}^2\right>^2-\frac{2}{V^4}\sum_i m_i^4.
	\end{align}
At criticality, \(U\) is independent of \(L\), while in the ordered phase \(U(L\to\infty)\to 1\), and in the disordered phase \(U(L\to\infty)\to 0\).
Note that the high order of \(U\) makes it sensitive to fluctuations, meaning it may not give reliable results in the brain samples.
Calculating this quantity (Fig.~\ref{fig:criticality-tests}(b)), we find that for the fly \(U\) decreases with \(L\), while in the mouse and human, \(U\) is roughly constant at smaller \(L\) and decreases at the largest sizes, indicating the brain may be slightly in the disordered phase. 
The shading in Fig.~\ref{fig:criticality-tests}(b) shows the region of the plot in which site percolation samples with occupancy \(0.95\leq p/p_c \leq 1.05\) fall, where \(p_c\) is the critical occupancy. 
If percolation can offer any guidance on the brain behavior, these bounds reinforce the idea that the brain is close to criticality.
Taken together, the spanning fraction and Binder cumulant measurements are inconclusive in determining on which side of the critical point the brain lies, although both suggest that brain structure is close to criticality.

\section{Scaling relations and universality}
We provide further support for the idea of structural criticality in the brain by examining the scaling properties of the critical exponents.
At criticality, the observed critical exponents are not independent and obey certain scaling relations, such as Eq.~\ref{eq:tau-gamma}. 
We can test the validity of another scaling relation in the brain using Fisher's identity \(\gamma/\nu = 2-\eta\)~\cite{Fisher1964}, by calculating \(\eta\) values from our \(\gamma/\nu\) estimates.
The results, listed in Tab.~\ref{tab:critical-exponents2}, are consistent with the \(\eta\) values calculated from \(C(r)\) (Tab.~\ref{tab:critical-exponents}), suggesting that the brain exponents obey standard scaling relations.

The observation that scaling relations hold between independently measured critical exponents is strong evidence that the structure of the brain has evolved to be at, or at least close to, a state of criticality as understood in physical systems. 
This means that the individually observed long-range patterns are not independent quantities.
Just like in critical physical systems, many other characteristics of brain structure could be measured when suitable data becomes available.
This would lead to a standard set of critical exponents in each brain, which would define a universality class.
Among these there are at most two or three independent exponents, while all others are related via scaling relations. 
The number of independent exponents depends on the validity of certain \textit{hyperscaling} relations involving the spatial dimension \(d\).

We examine whether hyperscaling holds in the brain by examining the hyperscaling relation 
	\begin{equation}
	\eta=2+d-2d_f.
	\label{eq:hyperscaling}
	\end{equation}
We introduce the ``hyperscaling violation exponent'' \(\theta\), which is zero if hyperscaling is valid~\cite{Aharony1976,Fernandez2011,Campbell2019}. 
For percolation, hyperscaling is obeyed (\(\theta=0\)) below the upper critical dimension \(d=6\), while \(\theta = d-6\) for \(d>6\). 
In general Eq.~\ref{eq:hyperscaling} becomes  \(\theta = 2 + d-2d_f-\eta\), or, by using Fisher's identity, \(\theta = d-2d_f+\gamma/\nu\).
The resulting \(\theta\) values, listed in Tab.~\ref{tab:critical-exponents2}, are consistent overall between brains and indicate that the brain violates hyperscaling with \(\theta\sim 1\).

Comparing our brain exponent estimates in Tabs.~\ref{tab:critical-exponents} and \ref{tab:critical-exponents2}, we observe that the exponent values obtained appear consistent between the three organisms studied.
In this comparison, we note that our error estimates are based only on statistical properties and do not account for biological variation or uncertainty due to data quality, meaning the true uncertainty is potentially much larger.
The apparent consistency we observe between the exponent values of different brains suggests that, to a first approximation, the brains of multiple organisms may be described by a single \textit{structural universality class}.

\section{Discussion}
Our results indicate that the brains of multiple organisms show signatures of being at or close to a structural phase transition and that the corresponding critical exponents are consistent between organisms. 
If these structural properties of the brain are indeed critical and universal between organisms, we would expect to observe consistent exponents in other organisms and brain regions.
However, with the studied datasets we have only been able to analyze a single partial brain region in three organisms, with the largest lengthscale probed being limited by the overall size of the brain samples.
We therefore cannot determine whether these measurements are sensitive to known structural variability between brain regions, as well as variations between individual organisms.
We also note that the statistical sampling techniques used here could not be conclusively applied to small nervous systems, and that since criticality is a large-scale phenomenon we anticipate that there would be a minimum brain size above which these features emerge.

Within the datasets, the incomplete proofreading of the human~\cite{human} and mouse~\cite{mouse} data may also impact results at large $L$, in addition to the neuron analysis discussed previously.
In particular, split and merge errors in the segmentation data mean that individual cells may be classified as multiple cells, or multiple cells may have a single segmentation ID, which we expect would have the most impact at the largest sample sizes we study.
The fact that these volumes are partial brain regions also impacts the segmentation due to cell fragments that would be connected outside of the sample being classified as separate cells within the volume. 
We explored this effect by relabeling segments within our 3D samples so that only regions of the same cell that were in physical contact within the sample were treated as the same segment. 
Our analysis (Fig.~S4) indicates that the long-range behavior is robust to this change, with results qualitatively similar to the original segmentation.

When comparing results between brains, it is important to consider the relative length scales involved.  
The entire fly brain is about the size of a dendritic arbor in a mammalian cortical pyramidal neuron~\cite{fly}.
As such, in the fly we are able to probe longer distances relative to the length scale of the entire brain than are accessible in the mouse and human.
In the mammalian samples, the available length scale may only start to probe the regime of true long-range behavior.
This could explain the FSS of the critical exponents in the mouse and human, where the largest sizes appear to have a different trend to the smaller sizes.
While the present analysis is limited by the physical length scale of each dataset, some results could be verified at larger scales in the future using full-scale reconstructions of select neurons, such as the Allen Institute single cell data in the human~\cite{AllenHuman} and mouse~\cite{AllenMouse}.
As more data becomes available, it will be possible to expand our analysis to study larger scales~\cite{flywire, Zheng2018}, further brain regions, additional organisms, including ongoing experiments in zebra fish~\cite{DeWeerdt2019}, and individual variation.
Overall, we believe that with increasing data coverage and quality, statistical physics tools will become more and more powerful in characterizing brain data.

The exponents and scaling relations in the brain can be used as a benchmark for comparing structural models with actual properties of the brain.
For example, recent physical network models, in which the nodes have $d_f\sim1.62$, have been shown to have network properties reminiscent of the brain~\cite{Pete2023}.
Further analysis could determine whether other exponents in this model are also consistent with those observed in the brains.
The validity of scaling relations and the universality in the structure of the brain also open up the possibility of designing generative physical models within the brain universality class.
These models could be used to predict further universal aspects of brain structure including standard critical exponents and universal features that go beyond these exponents, such as the gap-size exponent \(\zeta\) and universal amplitude ratios.
Generative models could serve as a baseline model of brain anatomy in a wide range of organisms, and could also be useful for making predictions about brain structure and dynamics even if they are agnostic towards certain biological details.

\begin{figure}
\centering
\includegraphics[width=.8\linewidth]{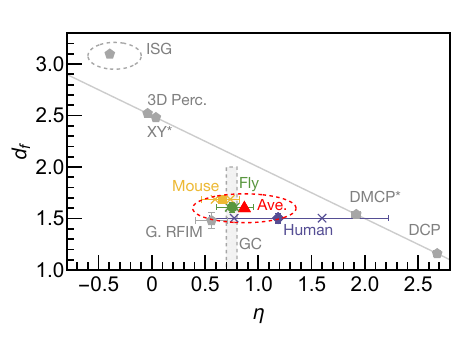}
\vskip-5mm
\caption{
Summary of \(\eta\) and \(d_f\) in the brain and a range of 3D physical systems.
Brain \(d_f\) values are the \(d_f^{(\mu)}\) values in Tab.~\ref{tab:critical-exponents}. The \(\eta\) values are the mean of those listed in Tabs.~\ref{tab:critical-exponents} and \ref{tab:critical-exponents2}, with \(\times\) indicating the individual values of the two \(\eta\) measurements and error bars showing the full range of uncertainty across the two values.
The Ave. point represents the mean exponent values for the brains, while the surrounding ellipse highlights the approximate region of the space into which the brain exponents fall.
The remaining points represent the following 3D systems: 
(3D Perc.) 3D Percolation~\cite{Xu2014}, 
(ISG) Ising Spin Glass (plotted point from values in Refs.~\cite{Katzgraber2006,Campbell2019}, ellipse axes represent the range of exponent values from studies listed in~\cite{Katzgraber2006}), 
(XY*) XY model~\cite{Campostrini2001}, 
(G. RFIM) Gaussian Random Field Ising Model~\cite{Middleton2002}, 
(DCP) critical point of the disordered quantum Ising model/Contact Process,~\cite{Kovacs2011a,Kovacs2011b},
(DMCP*) multicritical point of the disordered quantum Ising model/Contact Process~\cite{Kovacs2021},
(GC) Galaxy clustering~\cite{Coleman1992,Conde2015}. 
The plotted gray line indicates the hyperscaling relation \(\eta = 2+d-2d_f\), * indicates hyperscaling is assumed to hold.
}
\label{fig:exponent-plot}
\end{figure}

As a first consideration towards a generative brain model, we compare the brain critical exponents with exponents in other physical systems.
We highlight a broad range of 3D systems in Fig.~\ref{fig:exponent-plot}, focusing on \(\eta\) and \(d_f\), which we estimated directly.
Despite showcasing percolation as an example of structural criticality, it is unsurprising that this system, where random occupation of sites or bonds leads to clusters, provides a poor model for brain structure.
We observe that the brain exponents are close to those of the Gaussian Random Field Ising Model (G. RFIM)~\cite{Middleton2002}, which violates hyperscaling with \(\theta\sim 1.5\).
This model may therefore offer guidance on expected values of other brain exponents, such as $\nu_{\mathrm{G. RFIM}}\sim 1.4$,
although defining clusters in the G. RFIM analogous to the brain cell fragments remains an open challenge. 
It is intriguing that in the cosmic web, one of the few other systems where these exponents are known, the \(\eta\) and \(d_f\) ranges overlap with those of the brain~\cite{Coleman1992, Conde2015}.
However, it is unclear to what extent the universe could be interpreted as a critical system.

While the possibility of designing generative brain models makes structural criticality an appealing physical model for the brain, it is important to consider the biological relevance.
Universal features of the brain can be directly compared between organisms, and can therefore be used as guidance for instances where one organism may be used as a model for another. 
The nature of the relationship between structural criticality in the cellular structure of the brain and observations of critical behavior in neuronal avalanches~\cite{Beggs2003} remains an exciting direction to explore. 
Indeed, models of critical neuronal activity, which may give the brain optimal computing and processing properties~\cite{Beggs2008,Shew2013}, do not require complexity in neuronal structure~\cite{Gollo2013}.

Network aspects of the brain, which facilitate functional properties, are inevitably underpinned by the cellular structure.
The structure of the brain neural network is expected to be influenced by the wiring cost~\cite{Cajal1899,Ahn2006, Chen2006, Bassett2010} and functional constraints on the network~\cite{Chen2006}.
It has been shown that networks in \textit{C. Elegans}, large-scale human brain data, and \textit{Drosophila} display Rentian scaling~\cite{Bassett2010, fly}, which in the brain relates the number of connected neurons in a group with the number of connections out of the group through a power-law relation.
Analysis of the topological and physical Rent exponents in the network has shown that the wiring is efficient, but not minimized~\cite{Bassett2010}. 
Self-similarity in neuron structure has previously been proposed as a factor in balancing connectivity and cost~\cite{Wen2009,Chen2006,Smith2021}.
It is intriguing to consider that the brain may have evolved a structure close to criticality in order to aid formation of near-optimal network properties.
If the structure of the brain sat far into an ordered cluster phase, the structure would have many redundant long-range connections, or excess wiring, between different points.
Conversely, in the disordered phase there are no long-range connections, which are required to satisfy connectivity constraints. 
We therefore hypothesize that criticality in the structure may allow for the emergent near-optimal network properties by balancing geometric wiring costs with long-range connectivity requirements.
To understand how these anatomic and network properties come about in the brain, it would be insightful to study brain structure during development, pruning and learning as suitable data becomes available.



\begin{table}[ht]
       \caption{Summary of directly measured critical exponents in each brain.
    Numbers in parentheses give the measured uncertainty in the last digit(s) while numbers in square brackets give the difference between the largest \(L\) data point and the extrapolated value. 
    Superscripts indicate the method used to calculate the exponent. 
   For 3D percolation, the quoted number is based on numerical results from the literature, while the numbers in curly braces indicate the largest of the difference between our numerical result and the known result, and the uncertainty on our numerical result.}
    \centering
 	\begin{tabular}{cccccc}
    \toprule
       System & \(d_f^{(\mu)}\) & \(d_f^{(box)}\) &  \(\eta\) & \(\zeta\) & \(\gamma/\nu \)\\
        \hline
       Fly & \(1.61(5)[2]\) & \(1.42(1)\)&  \(0.8(2)[2]\) & \(1.79(5)[22]\) & 1.3(1)[1]\\
       Mouse  & \(1.69(4)[25]\) &  \(1.61(1)\) & \(0.6(1)[1]\) & \(1.9(1)[1]\) & 1.3(1)[1]\\
       Human & \(1.50(5)[29]\) & \(1.39(1)\) & \(1.6(6)[1]\) & \(2.1(5)[1]\) & 1.2(2)[1]\\
       Perc.\footnote{References~\cite{Wang2013,Kovacs2014Perc}}& \(2.52\{7\}\) & \(2.52\{11\}\) & \(-0.05\{3\}\) & \(2.00\{3\}\) & 2.05\{6\}\\
         \hline
         \hline
    \end{tabular}
    \label{tab:critical-exponents}
\end{table}

\begin{table}[ht]
    \centering
     \caption{Summary of indirectly measured critical exponents in each brain. 
    Superscripts in each column heading indicate the directly measured exponent in Tab.~\ref{tab:critical-exponents} used to calculate each value in the relevant (hyper)scaling relation, with \(d_f^{(\mu)}\) used for \(d_f\).
    The number given in parentheses gives the propagated uncertainty in the last digit(s).
     For 3D percolation, the quoted number is based on numerical results from the literature, while the numbers in curly braces indicate the largest of the difference between our numerical result and the known result, and the uncertainty on our numerical result.}
    \begin{tabular}{c c c c c c}
    \toprule
       System & \(\eta^{(\gamma/\nu)}\) & \(\tau^{(\gamma/\nu)}\) &\(\tau^{(\eta)}\) & \(\theta^{(\eta)}\) & \(\theta^{(\gamma/\nu)}\)\\
        \hline
       Fly & 0.7(1) & 2.2(1) &  2.2(1) & 1.0(2) & 1.0(1) \\
       Mouse  & 0.7(1) & 2.25(5) & 2.17(7) & 1.0(1) & 0.9(1) \\
       Human & 0.8(2)& 2.2(1) & 2.7(4) & 0.4(6) & 1.2(2) \\
       Perc.\footnote{References~\cite{Wang2013,Lorenz1998}} & \(-0.05\{6\}\)  & 2.189\{6\} & 2.189\{33\} & \(0.0\{2 \}\) & \(0.0\{1\}\)\\
         \hline
         \hline
    \end{tabular}
    \label{tab:critical-exponents2}
\end{table}

\section*{Acknowledgements}
We thank Anastasiya Salova and Sam Frank for assistance with accessing and filtering the brain data and for useful discussions. 
We also thank Bingjie Hao and Ruiting Xie for useful discussions.
This research was supported in part through the computational resources and staff contributions provided for the Quest high performance computing facility at Northwestern University which is jointly supported by the Office of the Provost, the Office for Research, and Northwestern University Information Technology.

\section*{Methods}
\textbf{Brain dataset sampling resolution -- }We utilize volumetric brain datasets for the human cortex~\cite{human}, mouse cortex~\cite{mouse}, and fruit fly (adult \textit{Drosophila melanogaster}) central brain~\cite{fly}.
The human data~\cite{human} consists of \(\sim1 \,\mathrm{mm}^3\) taken from the temporal lobe of the cerebral cortex.
This sample is wedge-shaped with maximum dimensions of  \(3 \times 2 \times 0.170 \,\mathrm{mm}^3\), imaged at \(4\times4\times33\,\mathrm{nm}^3\) resolution.
The mouse data~\cite{mouse} consists of \(\sim 1 \,\mathrm{mm}^3\) of visual cortex with dimensions of approximately \(1.3\times 0.87 \times 0.82 \,\mathrm{mm}^3\), imaged at a resolution of \(4\times4\times40\,\mathrm{nm}^3\).
The fly hemibrain dataset~\cite{fly} has approximate dimensions \(0.25 \times 0.25 \times 0.25 \,\mathrm{mm}^3\), imaged at a resolution of \(8\times8\times 8\,\mathrm{nm}^3\), and contains approximately half the volume of the fly central brain.

We sample the segmentation data in each brain, which gives the ID number of the cell present at each location within the volume. 
The segmentation data is available at a range of resolutions (mip values), with successively increasing mip values corresponding to an increase by a factor of two in the voxel dimension in the \(x\) and \(y\) directions. 
Due to the larger value of the voxel size in the \(z\) direction in the human and mouse datasets, the \(z\) voxel dimensions at the smallest sizes do not double in these brain samples until mip 3.

To choose the preferred sampling resolution, in Fig.~S3 we first examine how the mean number of segments in 100 randomly sampled slices of linear size \(L = 32.768\,\mu\mathrm{m}\) changes with mip value for each brain for all cell types (left panels) and neuron fragments  (right panels, discussed below).
For mip 4 and below (higher resolutions) \(>95\%\) of neuron segments (neuron identification discussed below) remain within the sample window during downsampling. 
At mip values higher than 4 we observe a sharp drop off in the proportion of segments remaining within the sample.

The results in Fig.~S3 suggest that sampling at mip 3 or below would give the least loss of data during downsampling. 
However, sampling at lower mip values means that larger sample sizes are required to study the same physical length scale.
We therefore choose to sample each dataset at mip 4, at which the largest sample size we can generate is dictated by the geometry of the brain samples. 
The sampling and analysis presented here require a computational time of over 4 CPU years, while sampling at lower mip values would require significantly more time. 
Computational limitations impact our ability to perform sampling and calculations at the largest sizes at lower mip values.
At mip 4, the voxel dimensions are 
\(128\times128\times132\,\mathrm{nm}^3\) for the human sample,
\(128\times128\times160\,\mathrm{nm}^3\) for the mouse, and
\(128\times128\times128\, \mathrm{nm}^3\) for the fly.

Tab.~S1 shows the value of a single data point that would contribute to the FSS calculations for each scaling exponent calculated in the main text for each brain sample at mip 3 and mip 4. 
The values listed for \(d_f^{(\mu)}\) are for \(L = 65.356\,\mu\mathrm{m}\), the largest 3D sample size in all three brain datasets (discussed below). 
For \(1+\eta\) and \(\zeta\) the values given are for \(L = 131.072\,\mu\mathrm{m}\), the largest physical size sample available in all three brains. 
In each case the value is given at each resolution for all segments, and neuron segments (discussed below). 
We observe that the datapoint values are consistent between mip 4 and mip 3, thereby justifying our choice to use mip 4.\\

\textbf{Generating brain samples from datasets -- }We generate three sets of samples from the segmentation data for each brain - 2D slices, 3D volumes, and 3D reconstructions of select neurons - using CloudVolume (\url{https://github.com/seung-lab/cloud-volume.git}).
The 2D slices are regions of size \(L\times L\) sampled from the \(x-y\) plane.
The origin of each sample, located at the top-left corner, is randomly selected while ensuring that the sample is contained within the brain volume. 
Entries in the resulting \(L\times L\) matrix contain the segmentation ID of the cell present at each position.

We generate 2000 2D samples for each brain at each \(L\) value.
In units of mip 4 voxels, we sample \(L = 2^n\) for integers \(4\leq n\leq 12\) (up to \(L = 524.288\,\mu\mathrm{m}\)) in the human and mouse, and  \(4\leq n \leq 10\) (up to \(L = 131.072\,\mu \mathrm{m}\)) for the fly.
The maximum \(L\) in each case is dictated by the brain sample geometry. 
We utilize 2D samples where possible due to brain sample geometries allowing a larger sample size in 2D than 3D. 

In 3D we generate 1000 samples at each size using the same overall procedure as in 2D.
In each case, we sample \(L = 2^n\) for integers \(4\leq n \leq 9\) at mip 4. 
The maximum sample size is again dictated by the brain sample geometry.
The largest sample sizes are \(65.536\times65.536\times67.584\,\mu\mathrm{m}^3\) in the human, \(65.536\times65.536\times81.92\,\mu\mathrm{m}^3\) in the mouse, and \(65.536\times 65.536 \times 65.536 \,\mu\mathrm{m}^3\) in the fly.
Note that \(L\) values quoted in microns for 3D samples are given in terms of their \(x\) and \(y\) dimensions.

We reconstruct select proofread neurons (discussed below) in 3D by scanning through the entire volume of each dataset and saving all positions associated with a given neuron.
In the fly we gathered the positions at mip-4 resolution, while in the mouse and human we used mip-5.
The higher choice of mip in the mouse and human is due to the larger volume of the datasets and individual neurons, compared to the fly, which makes the computation more demanding.\\

\textbf{Selecting dataset neurons --} Each dataset provides some level of cell-type identification to accompany the segmentation data.
In the fly, which is the most extensively proofread, there are 97 900 traced neurons, of which 22 212 have somas present and 21 739 are classified as uncropped. 
Uncropped neurons are defined in the release as having most arbors within the imaged volume, but the soma may not be present~\cite{fly}.
Here, we choose to use the 21 739 traced, uncropped neurons in all calculations where we filter to include only neurons.

In the mouse dataset, a prediction model based on nucleus data was used to identify cells as either neurons or not neurons~\cite{mouse-nucleus}. 
Filtering to keep only neurons for which the soma is contained within the volume gives 72 789 neurons.
These are the cells we count as neurons for calculations, except for box-counting. 
Of these neurons, 601 have undergone some level of comprehensive proofreading, including 78 that have undergone extensive proofreading of both their axons and dendrites.
For the 3D neuron reconstructions used in box-counting, we use the 78 neurons that have undergone extensive proofreading.

In the human dataset, we use the provided somas table for cell-type classification. 
Filtering this table to include only neurons gives 15 827 neurons, which we use for all neuron calculations except box-counting.
For box-counting, we use 3D reconstructions of the 104 fully proofread neurons released with the dataset.\\

\textbf{Generating critical percolation samples --} We generate samples for 3D critical site percolation on a lattice of size \(L^3\) with free boundaries.
Each site in the lattice is occupied with probability \(p\) and is otherwise unoccupied.
If two nearest neighbor sites are both occupied, they are considered part of the same cluster.
By setting \(p\) to the critical probability \(p_c=0.311608\)~\cite{Wang2013}, the resulting cluster structure displays structural criticality.

In 3D critical bond percolation, all sites in the lattice are occupied and nearest neighbor sites are connected if a bond is present between them.
We generate samples by occupying bonds at the critical probability \(p_c = 0.24881\)~\cite{Wang2013} and identify the resulting clusters of connected sites.

For results using 3D samples, we generate 1000 samples for sizes \(L=2^n\) for integers \(4\leq n \leq 9\), in line with the voxel size of the largest 3D brain samples.
For results using 2D samples, we generate 2000 3D percolation samples for sizes \(L=2^n\) for integers \(4\leq n \leq 10\) and perform calculations on a 2D slice through the center of each sample.
The largest size is the same as the largest sample size, in voxels, available in the fly at mip 4.\\

\textbf{Finite size scaling (FSS) --} FSS is used to extrapolate the values of critical exponents in the thermodynamic limit by extrapolating their \(L\)-dependence.
For a function \(X(x)\), an exponent \(a\) is defined by
	\begin{equation}
		a = \lim_{x\to\infty} \frac{\mathrm{ln}(X(x))}{\mathrm{ln}(x)}.
	\end{equation}
We examine the FSS of \(a\) by comparing points at size \(x\) and \(x/2\), where \(x\propto L\), to obtain a two-point estimate for \(a\), given by

\begin{equation}
	a(x) = \frac{\mathrm{ln}(X(x))-\mathrm{ln}(X(x/2))}{\mathrm{ln}(2)}.
\end{equation}
We estimate the value of \(a\) by examining the behavior of \(a(x)\) against \(1/L\) and extrapolating the size dependence as to \(L\to\infty\) (\(1/L\to 0\)).

Note that generally, the exponents scale with a combination of terms, including \(L^{-1}\) and  \(L^{-1/\nu}\), as well as higher order terms that are relevant only at small sizes. 
However, we observe that for the largest sizes, the FSS of the exponents is approximately linear on the \(1/L\) plots.
We therefore use linear fits through this approximately linear region to extrapolate exponent values.
Note that the value of \(\nu\) remains unknown in the brain, and estimating it reliably from FSS would require much smaller errors.  

\textit{FSS for \(d_f^{(\mu)}\) and \(\gamma/\nu\) --}
For the largest segment size \(\mu\sim L^{d_f}\) and the mean cluster size \(S\sim L^{\gamma/\nu}\), we use datapoints at size \(L\) and \(L/2\) for each estimate.
We extrapolate \(d_f^{(\mu)}\) using a linear fit through points corresponding to the largest three sizes. 
For \(\gamma/\nu\), the approximately linear regime includes only the largest two points for each brain and the largest three points for percolation.

\textit{FSS for \(\eta\) --}
For \(C(r)\sim r^{-d+2-\eta}\), we calculate two different two-point estimates for the exponent, taking points at \(x = L/8\) and \(x = L/16\).
The larger, \(x = L/8\), is the largest value at which the curve in samples of size \(L\) follows the curve in samples of size \(2L\). 
Points at \(x = L/8\) and \(x = L/16\) are highlighted in Fig.~\ref{fig:correlations}(b) for the largest \(L\), and points at \(x=L/8\) are highlighted for all \(L\) in Fig.~S1.
Plotted values in Fig.~\ref{fig:correlations}(b) (inset) show the mean of these two values and the error bars give the range. 
The extrapolated exponents are calculated with a linear fit through points for the largest three sizes.

\textit{FSS for \(\zeta\) --}
For \(n(s)\sim s^{-\zeta}\), we again calculate two different two-point estimates.
For the brains these are taken at \(x = L/2\) and \(x = L/4\), while for percolation we use \(x = L/8\) and \(x = L/16\).
These two points are highlighted on each curve in Fig.~\ref{fig:correlations}(c) for the largest \(L\), while points at the larger \(x\) values are highlighted for all \(L\) in Fig.~S2.
The plotted exponent values in Fig.~\ref{fig:correlations}(c) (inset) show the mean of these two estimates while the uncertainties indicate the range.
Extrapolated \(\zeta\) values include the largest three sizes in the linear fit.\\

\textbf{Box counting fractal dimension --} We implement box counting by overlaying 3D grids of side length \(L_b\) onto our fractal objects.
The box sizes are chosen to be approximately evenly spaced on a logarithmic scale.
We use at least 30 box sizes on each sample, with sizes ranging from \(L_b = 1\) voxel up to the size of the fractal object.
In each case, we align one corner of the grid with the minimum \((x,y,z)\) coordinates of the fractal object and count the number of grid cells \(N_b\) that contain part of the fractal for each \(L_b\).
We then estimate \(d_f^{(box)}\) from the slope of the linear region of the log-log plot.
Our box counting analysis is performed on the set of 3D reconstructions of select proofread neurons and on the largest clusters in the largest \(L\) percolation samples.
Using the full available extent of the proofread neurons allows us to access larger length scales for this calculation than are accessible with the 3D samples.\\

\textbf{Relation between \(\tau\), \(\gamma/\nu\) and \(d_f\) --} The Fisher exponent \(\tau\) is related to \(d_f\) through \(\tau = 1+ d/d_f\). 
When hyperscaling is valid, \(\eta = 2+d-2d_f\), which together with the Fisher relation \(\gamma/\nu = 2-\eta\) leads to \(\gamma/\nu = 2d_f -d\). 
We then use this to eliminate \(d\) in the relation for \(\tau\), leading to Eq.~\ref{eq:tau-gamma}.
When hyperscaling is violated, we need to substitute \(d\to d-\theta\) in all \(d\)-dependent relations. 
This does not change the resulting expression in Eq.~\ref{eq:tau-gamma}.


%

\end{document}


\title{Supplementary Material\\
Unveiling universal aspects of the cellular anatomy of the brain}

\author{Helen S. Ansell}
\affiliation{Department of Physics and Astronomy, Northwestern University, Evanston, Illinois 60208, USA}

\author{Istv\'{a}n A. Kov\'{a}cs} 
\email{Corresponding author: istvan.kovacs@northwestern.edu}
\affiliation{Department of Physics and Astronomy, Northwestern University, Evanston, Illinois 60208, USA}
\affiliation{Northwestern Institute on Complex Systems, Northwestern University, Evanston, Illinois 60208, USA}
\affiliation{Department of Engineering Sciences and Applied Mathematics, Northwestern University, Evanston, IL 60208}

\date{\today}

\maketitle

\renewcommand{\thefigure}{\arabic{figure}}
\makeatletter
\renewcommand{\fnum@figure}{FIG. S\thefigure}
\makeatother

\setcounter{figure}{0}

\renewcommand{\thetable}{\arabic{table}}
\makeatletter
\renewcommand{\fnum@table}{TABLE S\thetable}
\makeatother
\setcounter{table}{0}

\begin{figure*}
\centering
\includegraphics[width=0.8\textwidth]{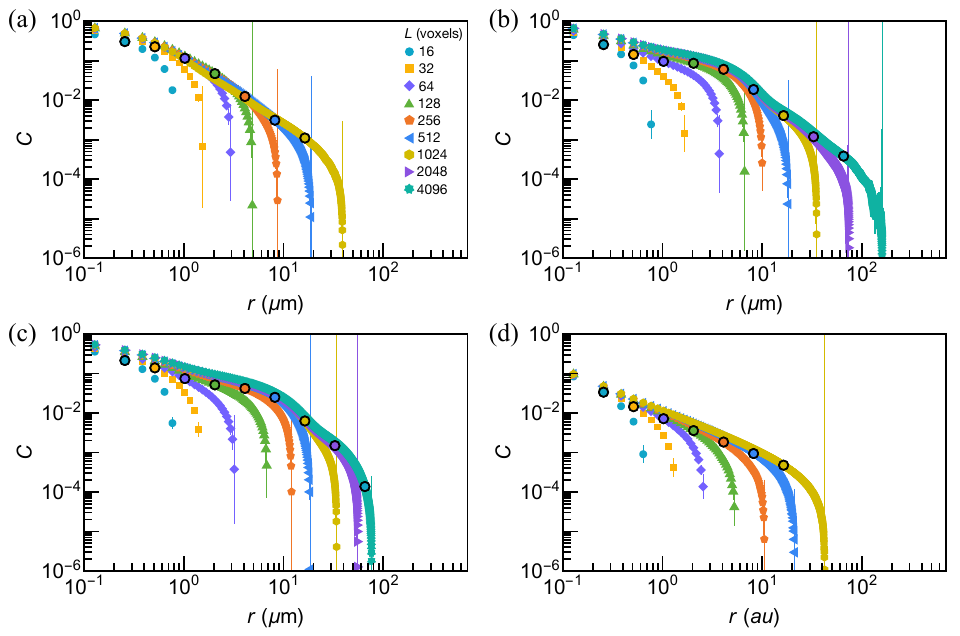}
\caption{
Calculated pair correlation function \(C(r)\) for different sample sizes (\(L\)).
Plots show \(C(r)\) for 
(a) the fruit fly brain, 
(b) the mouse brain, 
(c) the human brain, and 
(d) site percolation. 
The highlighted points at \(L/8\) are used along with points at \(L/16\) to calculate the two-point fits for the \(\eta\) exponent.  
}
\label{fig:2pt-size}
\end{figure*}

\begin{figure*}
\centering
\includegraphics[width=0.8\textwidth]{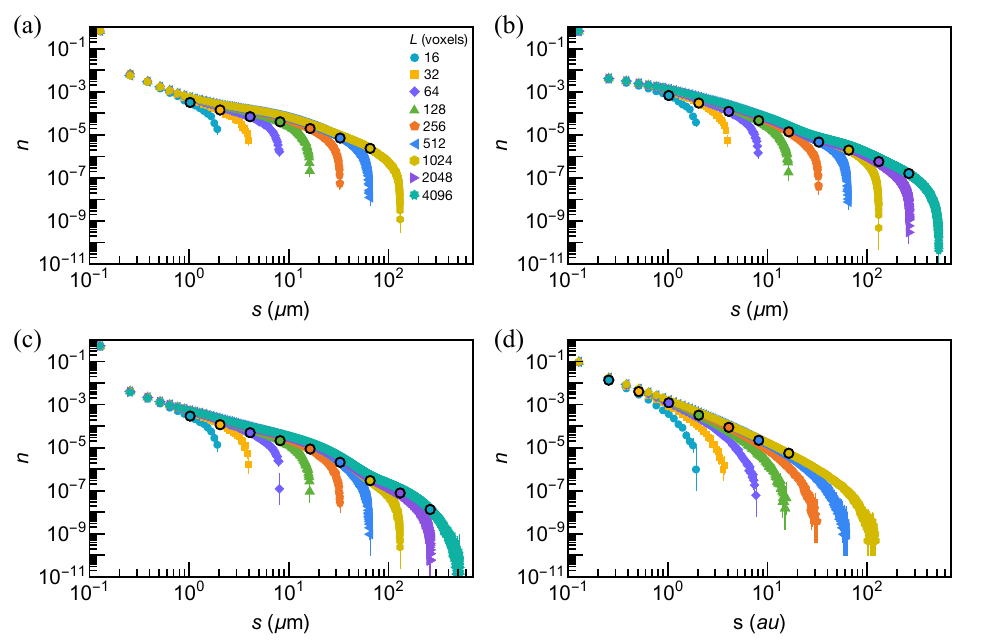}
\caption{
Gap size statistics \(n(s)\) for different sample sizes \(L\).
Plots show \(n(s)\) for 
(a) the fruit fly brain, 
(b) the mouse brain, 
(c) the human brain, and 
(d) site percolation. 
The highlighted points in (a-c) are at \(L/2\) while in (D) the highlighted points are at \(L/8\). 
These points, along with those at \(L/4\) and \(L/16\) respectively for the brains and percolation, are used to determine the \(\zeta\) exponent.
}
\label{fig:gap-size-size}
\end{figure*}

\newpage
\begin{figure*}
\centering
\includegraphics[width=0.8\textwidth]{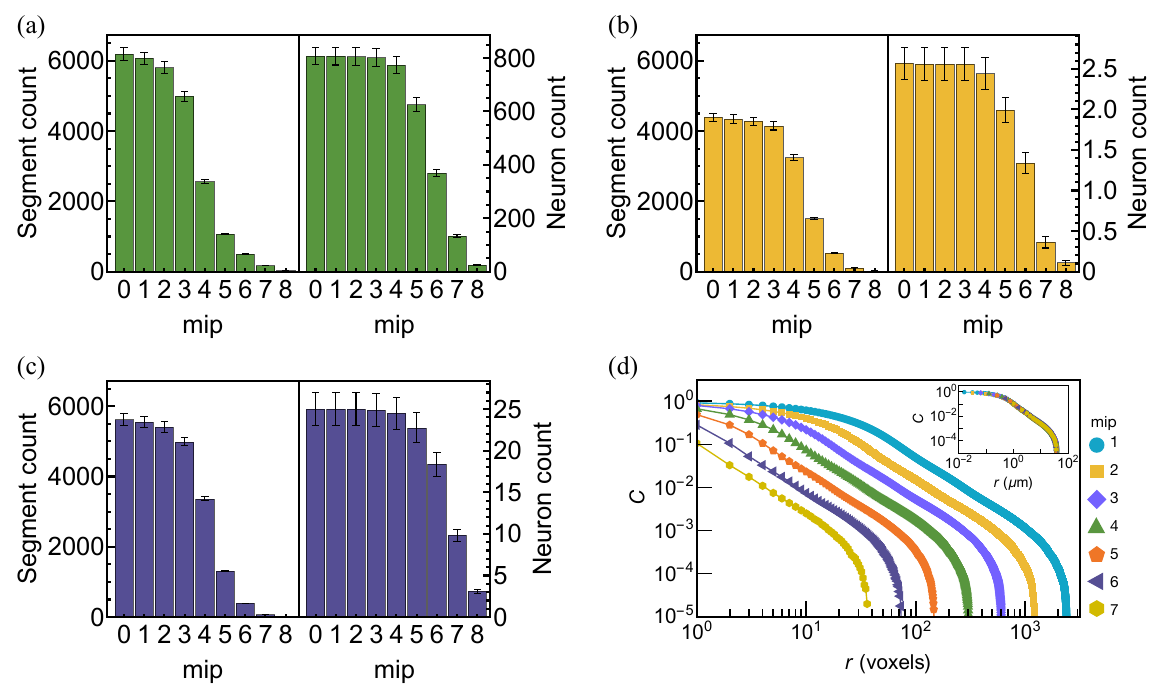}
\caption{Comparison of data sampled at different resolutions (mip values).
(a-c) Mean number of segments in 100 2D square samples of linear size \(L= 32.768\,\mu\mathrm{m}\), sampled at different mip values, for the (a) fly, (b) mouse, and (c) human datasets. The right panel of each figure shows the mean number of neuron segments in each sample. Note the different axis scales in the right panels of each plot.
These numbers help inform our decision to primarily sample at mip 4, although we verify our results are consistent with mip 3 in Tab.~S1.
(d) \(C(r)\) for 100 samples in the fly of size \(L= 131.072\,\mu\mathrm{m}\) at different mip values. 
(inset) Plotting \(r\) in \(\mu m\) instead of voxel units results in the curves collapsing, indicating consistent slopes between mip values.}
\label{fig:mip-comparison}
\end{figure*}

\newpage
\begin{figure*}
\centering
\includegraphics[width=0.8\textwidth]{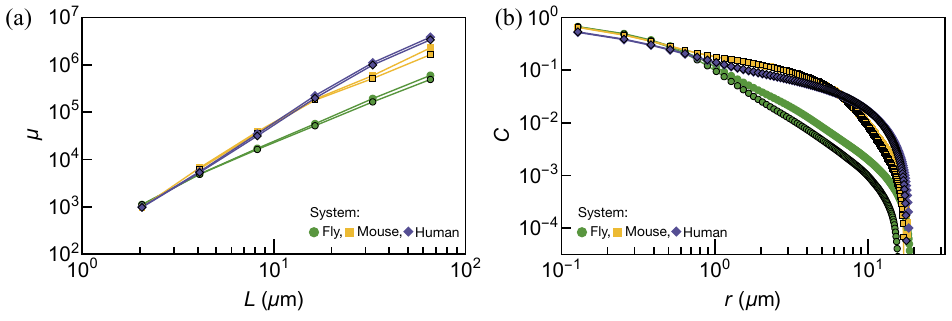}
\caption{Comparison between properties of the original segmentation and clusters derived from the segmentation based on connected components within a sample.
For each 3D sample, we analyzed each segment and assigned distinct cluster IDs to parts of the segment that were disconnected within the sample. We then performed calculations on these newly defined clusters to compare with properties of the original segmentation, as shown for 
(a) the largest cluster mass in 3D brain samples, and 
(b) \(C(r)\) for the largest \(L\) available in the 3D samples. In each plot markers with colored edges indicate the original segments while those with black edges show results for clusters after the connected components have been determined.}
\label{fig:hk}
\end{figure*}

\begin{table*}
    \centering
     \caption{Comparison of individual datapoints measured at mip 3 and mip 4 resolution.
    Each entry for an exponent corresponds to the value of a single datapoint used in FSS analysis. For \(d_f^{(\mu)}\), the points are for the largest available size \(L = 65.536\,\mu\mathrm{m}\), while for \(1+\eta\) and \(\zeta\) the points correspond to \(L = 131.072 \,\mu\mathrm{m}\), which is the largest size for which samples are available in all three brains. 
    }
    \begin{tabular}{|c |c |c| c| c |c |}
    \hline
     \multirow{1}{*}System   & Included segments & mip & \(d_f^{(\mu)}\)		&\(1+\eta\)	&\(\zeta\) \\\hline
        \multirow{4}{*}{Fly}   & \multirow{2}{*}{All} & 4 & 	\(1.63(3)\) & \(1.6(1)\)	& 	1.57(2)	\\\cline{3-6}
 	     &	 & 3 & \(1.68(3)\) & 1.6(1) & 1.55(3) \\ \cline{2-6}
     	    & 		\multirow{2}{*}{Neurons}   & 4 & 	\(1.65(3)\)  & \(1.6(1)\)  & 1.55(2) \\\cline{3-6}
	   & 			& 3& \(1.70(3)\) & \(1.5(1)\) & \(1.52(2)\) \\	\hline   
	\multirow{4}{*}{Mouse}   &\multirow{2}{*}{All} & 4 & \(1.94(3)\) & \(1.9(3)\) & 1.5(2) \\\cline{3-6}
 	    	& &	  3 &\(1.89(3)\) &  1.9(3)& 1.4(2) \\\cline{2-6} 
     	    & 	\multirow{2}{*}{Neurons}  & 4 & 	\(2.5(1)\)   & 2.4(7) &1.1(2)\\\cline{3-6}
	   & 				& 3& \(2.7(1)\) & 2.5(6)&1.83(2) \\\cline{3-6}		   
	   \hline
		\multirow{4}{*}{Human}   &\multirow{2}{*}{All}  & 4 & \(1.79(3)\)  &\(1.4(6)\) & 2.5(4)\\\cline{3-6}
 	     &	 & 3 & 1.78(3) & 1.5(6) & 1.98(2) \\ \cline{2-6}
     	    & 	\multirow{2}{*}{Neurons}   &  4 & 	\(1.78(3)\)  & \(1.7(9)\) & 0.85(6) \\\cline{3-6}
	   & 				& 3& \(1.83(5)\) & 1.8(9) & 1.0(1) \\		  
	\hline
    \end{tabular}
       \label{tab:critical-exponents-mip-comparison}
\end{table*}